# Cosmology of the Λ-term (vacuum component)

Vladimir Burdyuzha

Astro-Space Center, Lebedev Physical Institute, Russian Academy of Sciences, Profsoyuznaya str.84/32, Moscow 117997, Russia

burdyuzh@asc.rssi.ru

The vacuum component of the Universe is investigated in both quantum and classical regimes of its evolution. More than 78 orders of magnitude of the vacuum energy have been reduced in the quantum regime during $10^{-6}$ sec. Near 45 orders have been reduced in the classical regime during $4 \times 10^{17}$ sec. In the quantum regime phase transitions were more effective processes for vacuum energy reduction than production of new quantum states. The validity of evolution of the Universe's vacuum component is also presented. All the "crisis" 123 orders are, strangely enough, reduced in usual physical processes.

The dark energy (cosmological constant, Λ-term) problem might be solved only after introducing the holographic principle in physics [1] or more exactly after introduction of the entropic force [2]. Besides, it is definitely necessary to associate the dark energy (DE) with the vacuum energy of the Universe. It is practically an experimental fact [3]. In field equations the cosmological constant (Λ-term) was introduced by A. Einstein almost 100 years ago as the property of space:

$$G_{\mu\nu} + \Lambda g_{\mu\nu} = - 8\pi G_N T_{\mu\nu}. \qquad (1)$$

If we move the Λ-term to the right-hand side of this equation then it will be a form of energy that was named DE since we did not know the exact physical nature of this form of energy. The unsolved problem of DE even created a crisis of physics connected with the large difference (123 orders of magnitude) in the density of this form energy at the moment of the Universe creation (z=∞) with its present-day density (z=0).

$$\rho_{DE} \sim 2 \times 10^{76} (GeV)^4 \quad (\sim 0.5 \times 10^{94} \, g \, cm^{-3}) \quad for \quad z=\infty$$

$$\rho_{DE} \sim 10^{-47} (GeV)^4 \quad (\sim 0.7 \times 10^{-29}\ gcm^{-3}) \quad \text{for } z=0.$$

Here we want to show in which physical processes in our Universe the huge reduction of the vacuum energy (123 orders) took place. The Universe spent the vacuum energy for its expansion for organization of new quantum states all the time but in the initial period of its evolution other processes (phase transitions) were more effective for this reduction. It was a quantum regime, and phase transitions as a mechanism of huge reduction were already mentioned in [4]. Phase transitions produced condensates of quantum fields which compensated the positive vacuum energy of the Universe by 78 orders because the equation of state of any condensate is p= -ρ. A probable chain of phase transitions might be [5]:

P ⇒ $D_4 \times [SU(5)]_{SUSY}$ ⇒ $D_4 \times [U(1) \times SU(2) \times SU(3)]_{SUSY}$ ⇒

$10^{19}$ GeV $\qquad\qquad 10^{16}$ GeV $\qquad\qquad\qquad\qquad 10^{10} \sim 10^{5}$ GeV

⇒ $D_4 \times U(1) \times SU(2) \times SU(3)$ ⇒ $D_4 \times U(1) \times SU(3)$ ⇒ $D_4 \times U(1)$

$10^{10} \sim 10^{5}$ GeV $\qquad\qquad 10^{2}$ GeV $\qquad\qquad$ 0.15 GeV $\qquad\qquad$ ,

This chain might be more complicated (P→$E_6$ →O(10)→SU(5), for example). Above all, already on the Planck scale 3-dimensional topological defects of the gravitational vacuum condensate (worm holes) [6] have renormalized (diminished) the Λ-term:

$$\Lambda = \Lambda_0 - (\kappa \hbar^2 / 768\pi^2)\ c_3^2 \qquad (2)$$

here: $c_3$ is a constant in the expanding of a parametric function[6] ; κ is Einstein constant (in natural units). Note also that in the supersymmetric vacuum the energy density of a boson field ($\rho_{bos}$) and that of a fermion field ($\rho_{fer}$) had different signs and the total energy might be equal to zero:

$$<\rho_{bos}> = \infty; \quad <\rho_{fer}> = -\infty; \quad <\rho_{tot}> = 0. \qquad (3)$$

But as is known, later on the supersymmetry was broken and probably $<\rho>_{SS} \sim 10^{64}$ GeV$^4$ might be at that moment. The compensation was then necessary, obligatory. We cannot calculate the energy density of all condensates of our chain but the last two condensates of quantum fields in the framework of the Standard Model may be calculated exactly. They are named the Higgs condensate in the theory of the electro-weak interaction ($\rho_{EW}$) and the quark-gluon condensate in quantum chromodynamics ($\rho_{QCD}$). Therefore, as was shown in [6]:

$$\rho_{EW} = -m_H^2 m_W^2/2g^2 - (1/128\pi^2)(m_H^4 + 3m_Z^4 + 6m_W^4 - 12m_t^4) \qquad (4)$$

If Higgs mass is $m_H = 125$ GeV then we have $\rho_{EW} \approx -(100 \text{ GeV})^4$. If

$$\rho_{QCD} = -(b/32) < 0 \square (\alpha_s/\pi) G_{ik}{}^a G^{ik}{}_a \square 0 >, \qquad (5)$$

then we have $\rho_{QCD} \approx -(265 \text{ MeV})^4$ [7]. Only the quark-hadron phase transition 'quenches' near 10 orders, that is $\Delta\rho \sim (100/0.265)^4 \sim 2\times10^{10}$. Phase transitions have quenched more than $10^{78}$ orders of magnitude of the vacuum energy, $\Delta\rho \sim (M_{Pl}/M_{QCD})^4 = (1.22\times10^{19}/0.265)^4 \sim 4.5\times10^{78}$. Such a huge reduction of vacuum energy terminated at a moment when the Universe had an age of $\sim 10^{-6}$ sec only and, besides, at that moment the Universe also lost the chiral symmetry SU(3)$_L$ x SU(3)$_R$. The chiral QCD symmetry is not an exact one and pseudo-Goldstone bosons are a physical realization of this symmetry breaking. The spontaneous breaking of this symmetry leads to appearance of an octet of pseudoscalar Goldstone states in the spectrum of particles (π-mesons). In this process pi-mesons are excitations of the ground state and they definitely characterize this ground state (that is they characterize the QCD vacuum) [8]. Ya. Zel'dovich [9] 40 years ago attempted to calculate a nonzero vacuum energy of our Universe in terms of quantum fluctuations of particles as a high order effect. He inserted the mass of proton or electron into his formula but the result was not satisfactory. The situation changes if the average mass of pi-mesons ($m_\pi$ =138.04 MeV) is inserted in Ya. Zel'dovich's formula:

$$\Lambda = 8\pi G_N^2 m_\pi^6 h^{-4} \text{ cm}^{-2}; \quad \rho_\Lambda = G_N m_\pi^6 c^2 h^{-4} \text{ g cm}^{-3} \quad (6)$$

$$\Omega_\Lambda = \rho_\Lambda / \rho_{cr} = \Lambda c^2 / 3 H_0^2; \quad \rho_{cr} = 3 H_0^2 / 8\pi G_N. \quad (7)$$

If the Hubble constant is $H_0$ = 70.5 (kmsec$^{-1}$/ Mpc) then $\Omega_\Lambda$ ~ 0.73. This value of the vacuum component of the Universe might be correct if a condensate of the last phase transition (QCD condensate) was observed now. But the vacuum energy was evolving after the QCD phase transition and the magic coincidence of $\Omega_\Lambda$ with the modern value $\Omega_\Lambda$=0.73 is probably a strange result. For an energy of ~ 150 MeV (the end of the last phase transition) the vacuum energy stopped to drop quickly and later the vacuum energy decreased very slowly. Besides, by that moment (production of the QCD condensate at 1 μsec) the ratio of the Universe components ($\Omega_\Lambda$/ $\Omega_m$) had already hardened. But even at that moment there was still a large quantitative difference in the vacuum energy densities between the 'hardness' and the modern value:

$$\Delta\rho = (0.15 / 1.8\times10^{-12})^4 \sim 5 \times 10^{43}, \text{ if now } \rho_{DE} \sim (1.8 \times 10^{-12} \text{ GeV})^4.$$

Here, we are coming to an interesting point of our consideration. The Universe expands and new quantum states are produced for any matter components. The density of particles diluted as $1/R^3$ but the law of vacuum dilution was different ($1/R^2$). Besides, the vacuum energy is spent to produce new quantum states. Physical and mathematical basis of this statement can be proven. Physical grounds are the holographic principle [1] and T. Jacobson's idea [10] that gravitation on the macroscopic scale is a manifestation of vacuum thermodynamics. And also S. Hawking's idea [11] that the thermodynamics of a de Sitter universe is similar to the thermodynamics of a black hole in special coordinates. A mathematical ground is the Ostrogradsky-Gauss theorem introduced in cosmology by G. Smoot [12]. According to the holographic principle the physics of a 3D system can be described by a theory acting on its 2D boundary. J. Bekenstein [13] has shown that the entropy (number of quantum states) of a black hole is proportional to ¼ of its event horizon area expressed in the Planckian units. In cosmology, E. Verlinde's idea of an entropic force [2] is better to use for understanding although the ideas of holography in cosmology were not new. C. Balazs and I. Szapidi [14] obtained a formula for finding the Universe energy density in the holographic limit: $\rho \leq 3M_{Pl}^2/8\pi R^2$ where R is the apparent horizon of the Universe. Besides, important consequences of the holography take place: energy decreases linearly with increasing size of the Universe; the energy density decreases quadratically.

The authors [14] used the Fischler-Susskind cosmic holographic conjecture [15]: the entropy of the Universe (S) is limited by its "surface" measured in the Planck units: $S \leq \pi R^2 M_{Pl}^2$. It is easy to see that a relationship between the energy density and the number of quantum states of the Universe is established since the new quantum states have arisen due to expansion. Then in the holographic limit the energy density of the Universe connected with its entropy is:

$$\rho \sim 3 M_{Pl}^4 / 8 S \qquad (8)$$

Substituting to (8) the size of the Universe $R \sim 10^{28}$ cm we obtain the energy density of our Universe: $\rho \sim 10^{-57}$ (GeV)$^4$, if $M_{Pl} =1$. In the quantum regime of the Universe evolution the holographic conception does not work. The Universe entered the classical regime after the last phase transition at $E \sim 150$ MeV (of course, it was also transitional stage). If $R_{QCD} \sim 3 \times 10^4$ cm then $(R / R_{QCD})^2 \sim 10^{47}$. Finally, we have the simplest approximation formula for calculating the density of the vacuum energy $\rho_\Lambda (z)$ in the classical regime of the Universe evolution:

$$\rho_\Lambda (z) = (3/8) M_{pl}^4 [R_{QCD}/R(z)]^2 ; \text{ for } z=0 \quad \rho_\Lambda (0) = 0.375 \times 10^{-47} (GeV)^4 \qquad (9)$$

where R(z) can be calculated using "cosmological calculator" [16].

Thus: in the quantum regime of the Universe evolution during $10^{-6}$ sec the vacuum energy density decreased by 78 orders of magnitude from the Planckian value since in that epoch the positive vacuum energy density was affected by negative contributions producing condensates. By the end of phase transitions the vacuum energy density might be $\sim 10^{16}$ g cm$^{-3}$. In the classical regime of the Universe evolution the vacuum energy density decreased by 45 orders during $4 \times 10^{17}$ sec. Here the vacuum energy was spent on organization of new quantum states for expansion of the Universe. Probably, Bekenstein's thermodynamics of black holes may be a trace of the thermal nature of vacuum in the

Universe. If the DE is a vacuum energy then the quadratic dependence on z must necessarily occur. In [17] we have performed simple calculations of the vacuum energy evolution from z=0 till $z=10^{11}$. A physical basis for this was also given in [5]. At small redshifts (z<1) a smooth increase in the vacuum energy density takes place with growing z.  If this dependence is not detected in the nearest DE experiments [18] then the DE is a mix of the vacuum energy with another field (scalar as an example). Probably, the last doubts must disappear with respect to the vacuum energy evolution that is the vacuum energy of the Universe is a dynamical quantity $\rho_\Lambda$ (z). Besides, the recent measurements of time variations of the fine structure constant [19] and hypothetical time variations of other fundamental constants [20] will require revision of some physical foundations. An interesting question arises immediately. Is there a connection between time variations of fundamental constants and a time variation of the vacuum component of the Universe? Note, that cosmology with a time-dependent vacuum was already considered in [21]. Finally, practically an exact compensation of 123 orders of the vacuum energy of the Universe in the natural physical processes makes one believe these estimates and crisis of physics connected with the vacuum energy (cosmological constant, Λ-term) of the Universe thus may be overcome.

## References

1. G. t'Hooft , arXiv: hep-th/0003004.
2. E. Verlinde arXiv: 1001.0785; JHEP 1104(2011)29
3. E. Komatsu et al., arXiv: 0803.0547; Astrophys.J 700(2009)1097;
   E. Komatsu et al., arXiv: 1001.4538 v.3
4. R. Bousso , arXiv: 0708.4231; Gen.Rel.Grav. 40 (2008)607
5. V. Burdyuzha, Physics-Uspekhi 180 (2010) 439; Burdyuzha et al.,
   Phys. Rev. D 55 (1997) R7340
6. V. Burdyuzha, G. Vereshkov, Astrophys. Sp.Sci. 305 (2006) 235
7. L. Marochnik, D. Usikov, G.Vereshkov, arXiv: 0811.4484
8. V. Burdyuzha, Proceedings of the Symposium "PASCOS-98"
    Ed. P.Nath World Scientific 1999, p.101
9. Ya. Zel'dovich, Pis'ma JETP 6 (1967) 883; Physics-Uspekhi 95
    (1968) 209
10. T. Jacobson, Phys. Rev. Lett. 75 (1995) 260

11. S. Hawking, Commun. Math.Phys. 43 (1975) 199
12. G. Smoot, Int. J. Mod. Phys. 19 (2010) 2247
13. J. Bekenstain, Phys.Rev.D7 (1973) 2333
14. C. Balazs, I. Szapidi, arXiv: hep-th/0603133
15. W. Fischler, L. Susskind, arXiv: hep-th/9806039
16. N. Wright, Publ. Astron. Soc. Pacific. 118 (2006) 1711
17. V. Burdyuzha, arXiv: 1003.1025; Astron. Reports 56 (2012) 403
18. A. Albrecht et al., astro-ph/0609591; M. March et al., arXiv: 1101.1521, MNRAS 415 (2011) 143
19. T. Chiba, arXiv: 1111.0092; Prog. Theor. Phys. 126 (2011) 993
20. H. Fritzsch, J. Sola, arXiv: 1202.5097
21. J. Sola, arXiv: 1102.1815; J. Phys. Conf. Ser. 283 (2011) 012033